%% file: _PaperClusSpatCorr.tex
\documentclass{article}
\usepackage[
	backend=biber,
	datamodel=mydatamodel,
	style=alphabetic,
	natbib=true,
	url=false,
	doi=true,
	eprint=true,
	giveninits=true,
	isbn=false,
	maxbibnames=10,
]{biblatex}
\DeclareFieldFormat{mrnumber}{%
	MR\space
	\ifhyperref
	{\href{http://www.ams.org/mathscinet-getitem?mr=#1}{\nolinkurl{#1}}}
	{\nolinkurl{#1}}}
\DeclareFieldFormat{zbl}{%
	Zbl\space
	\ifhyperref
	{\href{https://zbmath.org/?q=an:#1}{\nolinkurl{#1}}}
	{\nolinkurl{#1}}}
\renewbibmacro*{doi+eprint+url}{%
	\iftoggle{bbx:doi}
	{\printfield{doi}}
	{}%
	\newunit\newblock
	\printfield{mrnumber}%
	\newunit\newblock
	\printfield{zbl}%
	\newunit\newblock
	\iftoggle{bbx:eprint}
	{\usebibmacro{eprint}}
	{}%
	\newunit\newblock
	\iftoggle{bbx:url}
	{\usebibmacro{url+urldate}}
	{}}
\usepackage[a4paper, margin=3cm]{geometry}
\usepackage{hyperref}
\usepackage{authblk}
\usepackage{parskip} % Layout with zero \parindent and non-zero \parsktip.
\input{packages}
\usepackage[capitalize,nameinlink,noabbrev]{cleveref}
\input{macros_general}
\newbox{\myorcidthanksbox}
\sbox{\myorcidthanksbox}{\large\includegraphics[height=1.8ex]{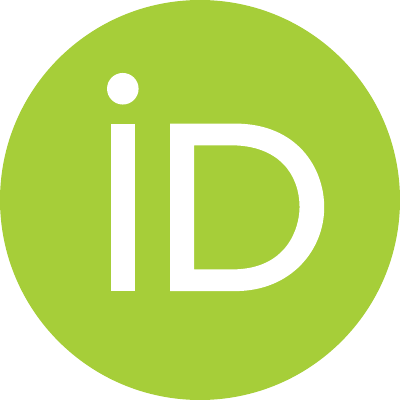}}
\newcommand{\orcidthanks}[1]{%
    \href{https://orcid.org/#1}{\raisebox{-0.5ex}{\usebox{\myorcidthanksbox}}\,#1}}
\def\plotPath{plots}%
\def\varyr{y}
\def\varcntry{c}
\title{Matters Arising: Spatial correlation in economic analysis of climate change\\{\small Arising from:  M.\ Kotz et al.\ Nature \url{https://doi.org/10.1038/s41586-024-07219-0} (2024) \cite{Kotz2024}.}}
\date{}
\author[1,2]{Christof Schötz\thanks{christof.schoetz@tum.de, \orcidthanks{0000-0003-3528-4544}}}
\affil[1]{Technical University of Munich, Germany; TUM School of Engineering and Design, Climate Center and Department of Aerospace \& Geodesy}
\affil[2]{Potsdam Institute for Climate Impact Research, Germany}
\begin{document}
\maketitle
Climate change poses substantial risks to the global economy \citep{ar6ch16}.
Kotz, Levermann and Wenz \citep{Kotz2024}, henceforth KLW, statistically analyzed economic and climate data, finding significant projected damages until mid-century and a divergence in outcomes between high- and low-emission scenarios thereafter. 
We find that their analysis underestimates uncertainty owing to large, unaccounted-for spatial correlations on the subnational level, rendering their results statistically insignificant when properly corrected. 
Thus, their study does not provide the robust empirical evidence needed to inform climate policy.

The economic impacts of climate change have been examined in multiple studies using panel data at the country--year level \citep{dell_temperature_2012, burke_global_2015, pretis_uncertain_2018, kahn_long-term_2021, krichene_social_2023}. A frequent approach is to regress economic growth rates on climate variables such as annual average temperature. Even when there is no true relationship, the regression fit will seemingly explain some of the variation in the economic growth owing to random fluctuations and entailed spurious correlations. Fortunately, the more information (in the form of data) we have, the better we are at distinguishing spurious
from true signals. 

More data can come from adding countries or extending the time span. KLW instead subdivide countries into subnational regions---about 20 per country on average---using the DOSE dataset \citep{dose}, thereby increasing the number of observations by an order of magnitude compared with country-level panels. However, there is no guarantee that this increase in data volume translates into a comparable increase in information.

To see why this is not necessarily the case, consider two extremes. If all subnational regions were fully independent, each would provide unique information, and the effective information would grow proportionally with the number of regions. However, if subnational regions of the same country were perfect copies of one another, no new information would be gained despite the larger dataset. 

The methods used by KLW implicitly rely on the first scenario---assuming that subnational regions contribute largely independent data. In reality, the situation more closely resembles the second scenario, as we demonstrate next by showing that empirical correlations between regions are large.

Uncertainty in regression models, such as those used by KLW, is encoded in the variances and correlations of the residuals of the model fit. An explanation of why we need to focus on the residuals here rather than, say, the predictors is given in \cref{sec:app:errorcorr}. As one cannot obtain meaningful uncertainty estimates under arbitrary correlation, one typically makes assumptions about which observations may be correlated and which are uncorrelated. In part, KLW account for temporal correlations, but never for any kind of spatial correlations. To examine this choice, we compute the average Pearson correlation coefficients $\bar\rho$ in different
clusters. We do this for the residuals in the largest model of KLW (ten lags for each variable), but the results hold qualitatively for reduced
models as well.

There is essentially no systematic correlation between residuals of the same region in arbitrarily different years ($\bar\rho = -0.03$) or consecutive years ($\bar\rho = 0.03$). Thus, the temporal correlation appears to be negligible. By contrast, there is a large positive correlation between different regions of the same country ($\bar\rho = 0.65$). As can be seen from \cref{tbl:corr}, this is also true for regions within the largest countries, but not in general for regions of different countries. Close regions in different countries have a small positive correlation on average, but less than regions in different countries in the European Union (EU).

\begin{table}[!h]
    \setlength{\tabcolsep}{4.5pt}
    \begin{tabular}{c|c|c|c|c|c|c|c|c|c}
        &&\multicolumn{3}{c}{Correlation coefficient}&\multicolumn{5}{c}{Correlation accounted for in clustering}
        \\
        Kind & 
        Group & 
        $\bar \rho$ & 
        Q25 & Q75 &
        Region &
        \multicolumn{1}{|p{0.9cm}|}{\centering Region --Year}&
        Country &
        \multicolumn{1}{|p{1.1cm}|}{\centering Country --Year}&
        Year \\
        \hline
        temp. & all                  & $-0.03$ & $-0.16$ & $0.08$ & yes & no & yes & no & no \\
        temp. & consecutive          & $0.03$ & $-0.12$ & $0.13$ & yes & no & yes & no & no \\
        spat. & all                  & $0.00$ & $-0.25$ & $0.24$ & no & no & partial & partial & yes \\
        spat. & same country (c.)        & $0.65$ & $0.55$ & $0.87$ & no & no & yes & yes & yes \\
        spat. & different c.      & $-0.01$ & $-0.25$ & $0.23$ & no & no & no & no & yes \\
        spat. & diff. EU28 c.     & $0.30$ & $0.11$ & $0.50$ & no & no & no & no & yes \\
        spat. & diff. EU'95 c.    & $0.36$ & $0.10$ & $0.64$ & no & no & no & no & yes \\
        spat. & $<$1000km, same c.  & $0.65$ & $0.54$ & $0.88$ & no & no & yes & yes & yes \\
        spat. & $<$1000km, diff. c. & $0.17$ & $-0.06$ & $0.44$ & no & no & no & no & yes \\
        spat. & $>$1000km, same c.  & $0.66$ & $0.57$ & $0.82$ & no & no & yes & yes & yes \\
        spat. & $>$1000km, diff. c. & $-0.02$ & $-0.26$ & $0.21$ & no & no & no & no & yes \\
        spat. & Russia               & $0.74$ & $0.67$ & $0.84$ & no & no & yes & yes & yes \\
        spat. & Canada               & $0.45$ & $0.22$ & $0.62$ & no & no & yes & yes & yes \\
        spat. & China                & $0.79$ & $0.73$ & $0.87$ & no & no & yes & yes & yes \\
        spat. & USA                  & $0.76$ & $0.72$ & $0.89$ & no & no & yes & yes & yes \\
        spat. & Brazil               & $0.79$ & $0.72$ & $0.87$ & no & no & yes & yes & yes \\
    \end{tabular}
    \caption{\textbf{Pearson correlation coefficients in different groups aggregated as mean $\bar\rho$, first quartile Q25 and third quartile Q75.} For the spatial kind, we have for each region one sequence of residuals indexed by year and calculate the correlation between the sequences. For the temporal kind, we have for each year one sequence of residuals indexed by the region and calculate the correlation between the sequences. We compute the mean and the first and third quartiles for the distribution of correlations of all pairs of different indices within the given group. The five rightmost columns show which clustering scheme (used for standard errors, cross-validation, and bootstrap) accounts for the correlations within each group. The EU28 group contains the EU countries between 2013 and 2020; the EU'95 group contains the EU countries between 1995 and 2004. Groups marked with $<$1000km ($>$1000km) contain only pairs of regions whose centroids are less (more) than 1000km apart.}\label{tbl:corr}
\end{table}

Regions in the same country (or in the same economic zone, such as the EU) typically have strong economic interdependencies that lead to highly correlated economic growth paths. This dependence is also manifested in the residuals of the regression model with climatic predictors, as the climate variables have low explanatory power (model without climatic predictors, $R^2 = 0.255$; model with all climate predictors with 10 lags each, $R^2 = 0.291$).

Having established the presence of spatial correlations in KLW’s analysis, we note that this issue has been previously recognized and addressed in the climate econometric literature \citep{Schlenker09PNAS,Hsiang10PNAS} with different solutions available \citep{Auffhammer2013}. However, KLW neglect spatial correlations, which are relevant in three parts of their work: the uncertainty of regression coefficients, model selection and the uncertainty of future damage projections. We show that KLW’s analysis can be corrected using methods they already apply.

First consider the uncertainty in the regression coefficients. A standard approach for valid inference under correlated data is to use clustered standard errors \citep{clusteredSe}: all observations are grouped into clusters; observations from different clusters are assumed to be uncorrelated; but correlations within a cluster are accounted for in the uncertainty estimate of the regression coefficients.

\begin{figure}
    \includegraphics[width=\textwidth]{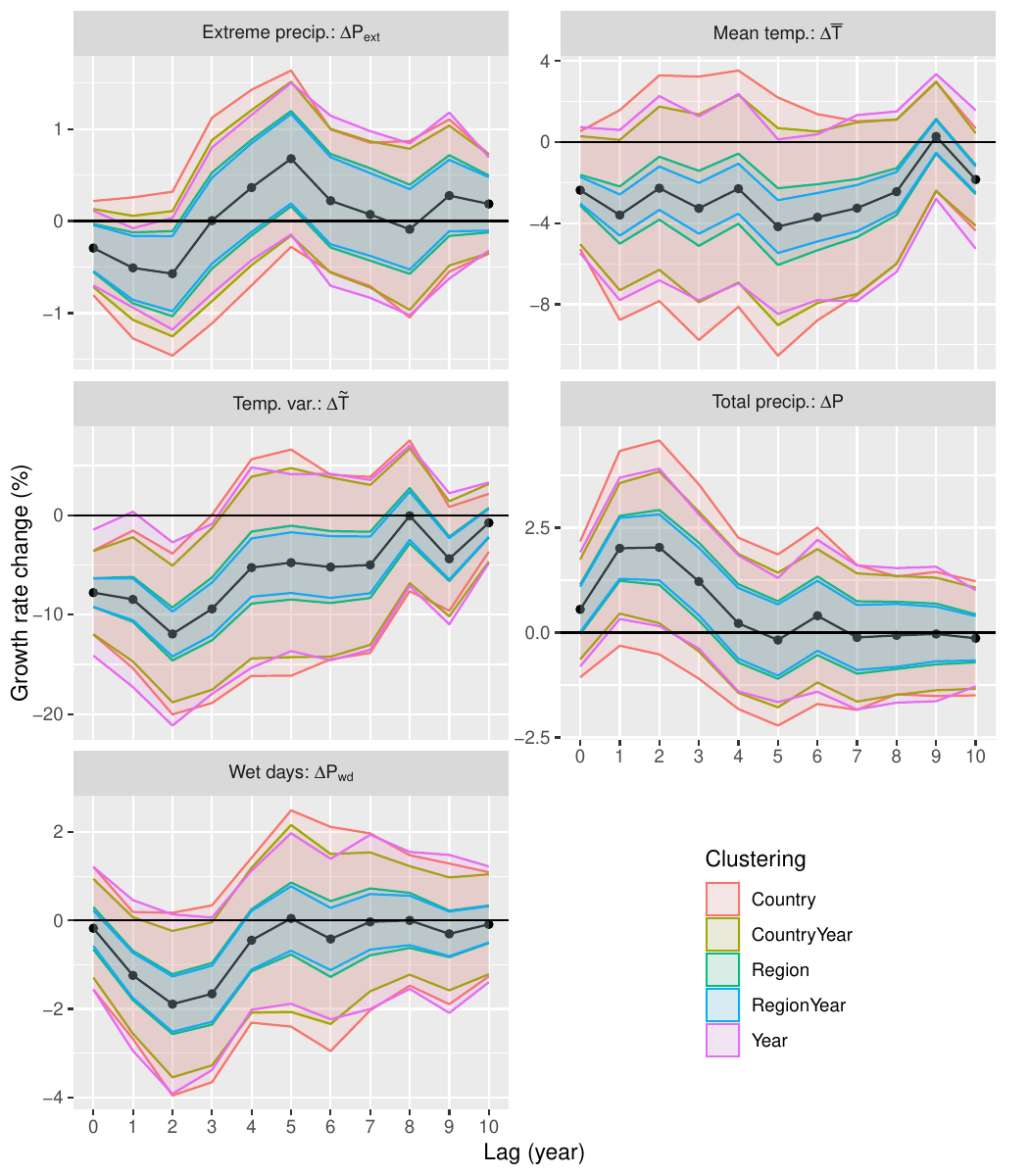}
    \caption{
        \textbf{Effects of different terms in KLW's full regression model.} A term consists of a variable in first difference form and its interaction with a moderator variable. Shaded areas show 95\% confidence intervals, computed using clustered standard errors with different clustering schemes. Clustering by \textit{region} reproduces the orange curves of KLW's Extended Data Fig.\ 1, where the moderator variable is set to its overall median.}
    \label{fig:significance}
\end{figure}

KLW use clustered standard errors. However, they define clusters as regions, meaning that different regions are assumed to be uncorrelated and the correlation between different years is taken into account. With the correlation analysis above, this does not seem reasonable. Using a clustering scheme that accounts for the strong spatial correlations, such as \textit{country–year}, we obtain uncertainty estimates in \cref{fig:significance} that are less biased.

As \textit{country–year} clustering ignores correlations between regions of different countries (such as within the EU), the results may still be overconfident. Other approaches, such as clustering by \textit{year}, account for these correlations, but reduce the number of clusters, making the uncertainty estimates themselves less reliable.

Second, we note that the underestimated uncertainties in KLW strongly impact the justification of their chosen model, in particular, the number of lag years. In KLW’s regression model, changes in climate variables can influence economic growth over multiple years. Although KLW’s choice of the maximum number of such lag years is based on a significance analysis for terms of variables and on the Akaike information criterion (AIC) \citep{aic} and the Bayesian information criterion (BIC) \citep{bic}, it
does not follow a fully formal procedure.

As \cref{fig:significance} shows, the corrected significance analysis discourages the use of many lag years and even shows that the most important predictor for KLW, annual mean temperature, has no significant coefficients.

Model selection via the information criteria AIC and BIC as applied by KLW assumes that the residuals are uncorrelated, which they are not, as shown above. We perform a simplified correction in \cref{sec:app:ic}. The results point to the trivial model without climate variables being preferred. Another alternative using cross-validation is shown in \cref{sec:app:cv}, which also discourages the use of most climate variables and lags.

Third, the uncertainties in KLW’s projected future damages are underestimated. KLW use a block bootstrap \citep{clusteredBoot} approach, where clusters of data (also called blocks) are formed to account for correlation within the blocks, whereas different blocks are assumed to be independent. Again, the authors account for temporal correlation and discard all spatial correlation by clustering by region.

\begin{figure}
    \begin{center}
        \includegraphics[height=0.9\textheight]{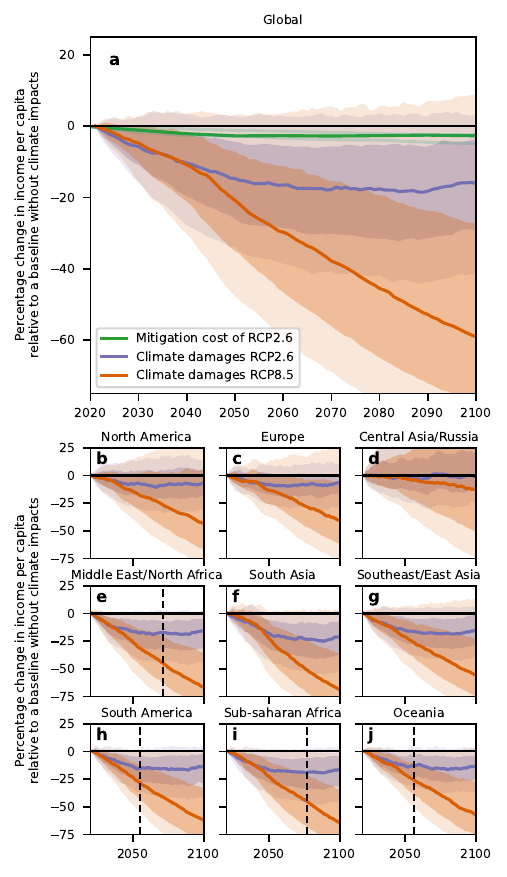}
    \end{center}
    \caption{
        \textbf{Reproduction of KLW's Figure 1 with bootstrap clustering by country--year.} Solid lines show median income loss; the shaded areas show 65 \% and 90\% confidence intervals, respectively. Vertical dashed lines indicate the first year in which the two emissions scenarios can be distinguished at the 5\% significance level. Missing dashed vertical lines in \textbf{a}, \textbf{b}, \textbf{c}, \textbf{d}, \textbf{f}, and \textbf{g} indicate that this distinction is not possible until at least 2100.
    }\label{fig:projection}
\end{figure}

To illustrate the effect of clustered spatial correlations, we use the main model specification by KLW (although corrected model selection procedures discourage this) and apply a block bootstrap with clustering by year. We reproduce KLW’s Figure 1 with the corrected version of the bootstrap in our \cref{fig:projection}. It shows much higher uncertainties and that the first year of discernible damages is shifted from 2049 to beyond the 2100 time horizon. This is also true for other clustering schemes that
account for correlations within countries, as shown in \cref{fig:gid0,fig:gid0Year,fig:gid1,fig:gid1Year} in \cref{sec:app:fig1}.

\clearpage

\subsection*{Data availability}
The results are based on the code and data used by KLW, which are publicly available via Zenodo at \url{https://doi.org/10.5281/zenodo.10562951}.

\subsection*{Code availability}
All code newly produced for this article is publicly available via \url{https://github.com/chroetz/ClusSpatCorr}.

\printbibliography

\clearpage

\begin{appendix}
	\section{Error Correlation}\label{sec:app:errorcorr}
	
	We explain why the choice of clustering depends on the correlations of the residuals in the regression model.
	
	\subsection{Error Correlation}
	The model used by KLW is an instance of a linear regression model: $Y_i = x_i\tr \beta + \varepsilon_i$, $i=1,\dots, n$, with target $Y_i \in\R$, predictor vector $x_i \in \R^p$, unknown parameter vector $\beta\in\R^p$, and error variable $\varepsilon_i\in\R$. Confidence intervals for the least squares estimate $\hat\beta$ are a function of the variance of $\hat\beta$, which in turn is a function of the $x_i$ and the covariance matrix $\Sigma\in\R^{n \times n}$ of the $\varepsilon_i$.
	
	To be precise, writing $X \in \R^{n\times p}$ the collection of predictor vectors as rows, and $Y \in \R^n$ the collection of targets in one vectors, we have $\hat\beta = (X\tr X)^{-1} X\tr Y$ and for the covariance matrix $\mathbf{Cov}(\hat\beta) = (X\tr X)^{-1} X\tr \Sigma X (X\tr X)^{-1}$.
	
	The values $x_i$ (and therefore $X$) are observed, but the covariance matrix $\Sigma$ is unknown and has to be estimated---at least indirectly. This is done using the residuals $r_i = Y_i - x_i\tr\hat\beta$ and a structural assumption on $\Sigma$, e.g.,  $\Sigma = \sigma^2 I_n$ for $\sigma\in\R_{\geq0}$ and the identity matrix $I_n\in\R^{n\times n}$ when we assume independent observations with homoscedastic noise. 
	
	Clustered standard errors \citep{clusteredSe} entail another kind of structural assumptions: Observations within a cluster are allowed to be arbitrarily correlated, corresponding to unknown arbitrary entries in $\Sigma$; but observations from different clusters are assumed to be uncorrelated, corresponding to $0$-entries in $\Sigma$. This means that $\Sigma$ has a block diagonal structure (if entries are ordered so that observations of the same cluster are consecutive).
	
	Thus, for identifying the correct way of calculating uncertainty (i.e., finding a meaningful structural assumption on $\Sigma$), the correlations and variances of the error variable $\varepsilon_i$ are the only thing that matters. While the values of the predictors influence the covariance matrix of the estimator $\hat\beta$, correlations of the predictor values have no influence on which structural assumption should be made on $\Sigma$, in particular, which clustering scheme should be used. This is also shown in the simulation study below.
	
	\subsection{Correlation Analysis of Residuals}
	
	A data-driven way of finding correlations of the error variable is a correlation analysis of the residuals. One should note that if error variables are perfectly uncorrelated, the residuals will still show some spurious empirical correlations. But if the number of samples is large enough and empirical correlations are averaged over enough observations, true correlations can reliably be distinguished from spurious ones.
	
	In our correlations analysis in the main text, the reported mean correlation for subnational regions of the same country is a mean over 26545 correlations of pairs of regions each estimated from a time series of on average 21 years. Given these large numbers, the high mean correlation value of $0.65$ and the high lower quartile value of $0.55$, we can consider this correlation to be relevant.
	
	\subsection{Simulation Study}
	
	In a simple simulation study of a standard linear regression model (\url{https://github.com/chroetz/ClusSpatCorr/blob/main/99_SimulationStudy_Correlation.R}), we show that correct clustering can be inferred from correlations of the residuals but not from the correlations of the predictor.
	We simulate a panel of 10 regions and 10 years. We randomly create a predictor variable with high temporal but low spatial correlations. We create the target variable as a linear function of the predictor variable plus noise sampled so that it exhibits high spatial correlation but low temporal correlation. 
	We apply different clustered standard error with clustering by region (accounting for temporal correlation) and clustering by year (accounting for spatial correlation). We repeat the experiment 1000 times.
	Clustering by region does not yield valid uncertainty estimates, but clustering by year---as suggested by empirical correlations in the residuals---does.
	
	In a second simulation study (\url{https://github.com/chroetz/ClusSpatCorr/blob/main/99_SimulationStudy_Bias.R}), we demonstrate that the clustered standard error estimator is unbiased---or at least exhibits only lower-order bias---when errors are independently distributed. This implies that applying clustering unnecessarily in such settings does neither lead to systematically too conservative or systematically too low uncertainty estimates. Instead, it yields estimates of lower quality than those from a more appropriate, non-clustered estimator, with deviations from the true value occurring in random directions.
	
	\subsection{Other Models}
	
	There are settings different from the linear regression model, where residual correlations may not be exclusively decisive for clustering choices.
	\citet{Abadie22} consider are binary treatment effects model, where they assume that observations or clusters of observations are sampled independently from a finite total population. They introduce estimates that---in this setting---better capture the true uncertainties compared to clustered standard errors. They even warn against relying on correlations in residuals for clustering choices. While this may be a valid point in their setting, it does not extend to linear regression models for climate econometric panels, for at least one (and likely more) important reason: neither countries, subnational regions, nor years in such datasets can be considered independent random samples from a broader population. In particular, the years are typically consecutive, not randomly drawn.

	\clearpage
	\section{Model Selection}\label{sec:app:cv}
	
\begin{figure}
		\includegraphics[width=\textwidth]{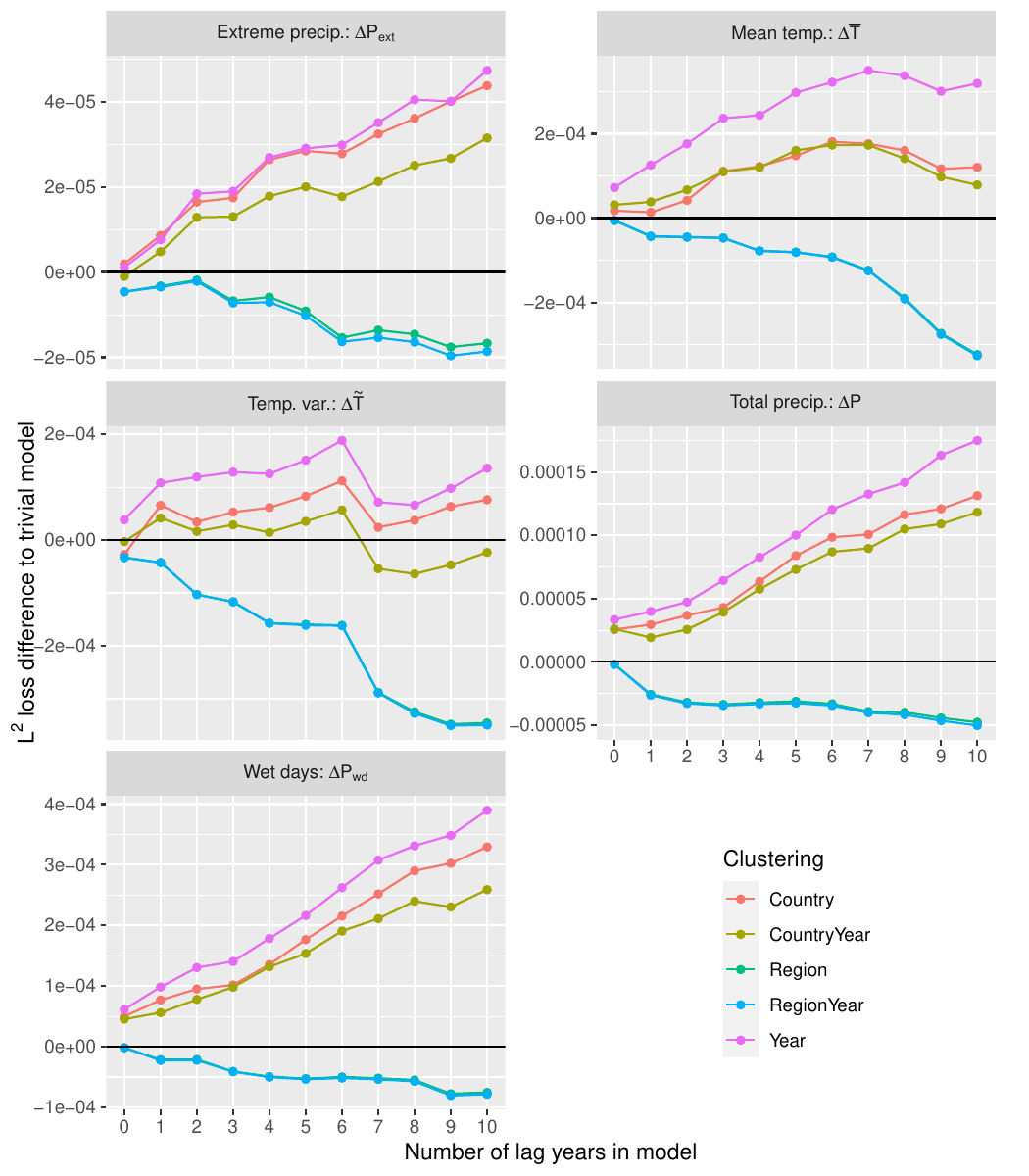}
		\caption{
			Change in cross-validated $L^2$-loss when adding a single term with lags to the trivial model (no climate variables) using different clustering. See \cref{fig:cvBack} for a similar plot with terms removed from the full model, which is more comparable to KLW's Extended Data Fig.\ 2. When using a clustering that ignores correlations between regions of the same country (\textit{Region}, \textit{RegionYear}), larger models are preferred. When this correlation is taken into account (\textit{Year}, \textit{Country}, \textit{CountryYear}), smaller models (often the trivial one) are preferred.
		}\label{fig:cv}
	\end{figure}

\begin{figure}
		\includegraphics[width=\textwidth]{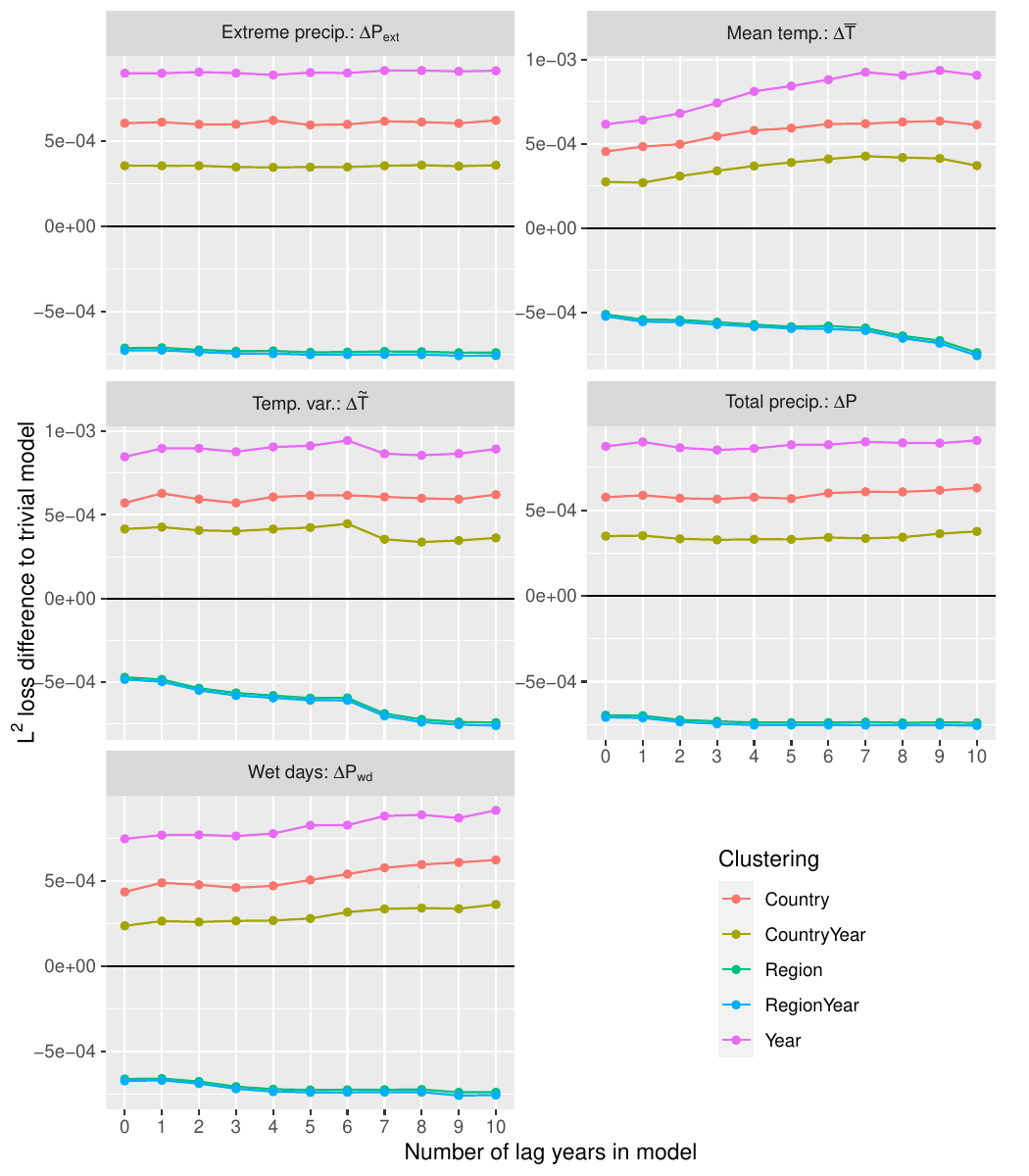}
		\caption{
			Change in cross-validated $L^2$-loss when removing lags of a single term from the full model with 10 lags per term using different clustering schemes. When using a clustering that ignores correlations between regions of the same country (\textit{Region}, \textit{RegionYear}), larger models are preferred. If this correlation is accounted for (\textit{Year}, \textit{Country}, \textit{CountryYear}), smaller models are preferred and the trivial model has the smallest loss.
		}\label{fig:cvBack}
	\end{figure}

	A valid alternative to information criteria for model selection, which can account for spatial correlations, is cross-validation. See \cite{Newell2021} for an application in a similar context (but note that these authors may not be able to find the best choice of control variables as they claim in their article). As with clustered standard errors, one can account for correlations in cross-validation by assigning observations to clusters \citep{clusteredCv}, which allows for correlations within clusters, but assumes independence between clusters. In each split of the data into training and validation set, all observations of the same cluster are assigned to the same set. We test several clustering schemes. Those that account for correlations of different regions of the same country discourage the use of most (or even all, in the case of clustering by year) terms of the main model specification by KLW, see \autoref{fig:cv} and \autoref{fig:cvBack}.
	
	\clearpage
	\section{Adjusted Information Criteria}\label{sec:app:ic}
	
	Both information criteria, AIC and BIC, are of the form
	\begin{equation}
		\gamma(n) k - 2 \log (\hat L)
	\end{equation}
	with $\gamma(n) = 2$ for AIC and $\gamma(n) = \log(n)$ for BIC, where $k$ is the number of estimated parameters and $\hat L$ is the maximized value of the likelihood function of the statistical model. If centered Gaussian errors are assumed, $\hat L$ becomes 
	\begin{equation}\label{eq:gauss}
		- \frac{n}2 \log(2\pi) - \frac12 \log(\det(\Sigma)) - \frac12 r\tr \Sigma^{-1} r,
	\end{equation}
	where $r\in\mathbb R^n$ denotes the residual vector and $\Sigma\in\mathbb R^{n\times n}$ the covariance matrix of the Gaussian distribution. If the residuals are assumed to be independent and identically distributed (iid), $\hat L$ further simplifies to
	\begin{equation}\label{eq:simplell}
		-\frac{n}2 \left(\log(2\pi) + \log(\sigma^2) + 1\right),
	\end{equation}
	where $\sigma^2 = \frac1n \sum_{i=1}^n r_i^2$ and $r_i$ is the residual of the $i$-th observation.
	KLW use the log-likelihood in the form of \eqref{eq:simplell}. By doing so, they assume uncorrelated residuals. We here present a modification of the standard form of the information criteria that accounts for correlations.
	
	We keep the Gaussian model, but adapt the covariance structure to better reflect the dependencies in the data. We assume that each country-year combination forms a cluster, different clusters are independent, and inside the same cluster different observations have covariance $\rho$ (same across all clusters). As in the iid case, we let $\sigma^2$ be the (constant) variance of each observation. Let $n_{\varyr, \varcntry}$ be the number of observations (observed regions) for country $\varcntry$ and year $\varyr$ and $\Sigma_{\varyr, \varcntry} \in \mathbb R^{n_{\varyr, \varcntry} \times n_{\varyr, \varcntry}}$ the respective covariance matrix, which is given by
	\begin{equation}
		\Sigma_{\varyr, \varcntry} := 
		\begin{pmatrix} 
			\sigma^2 & \rho & \rho & \hdots & \rho\\
			\rho & \sigma^2 & \rho & \hdots & \rho\\
			\rho & \rho & \sigma^2 &   & \vdots\\
			\vdots & \vdots & & \ddots & \rho\\
			\rho&  \rho & \dots & \rho & \sigma^2\\
		\end{pmatrix}.
	\end{equation}
	Then we set $n := \sum_{\varyr, \varcntry} n_{\varyr, \varcntry}$ and $\Sigma\in\mathbb R^{n\times n}$ is the block diagonal matrix of all $\Sigma_{\varyr, \varcntry}$ blocks.
	For convenience, we define $a:=\sigma^2 - \rho$. Then, we can calculate the relevant terms of the log likelihood in \eqref{eq:gauss} as follows:
	With the Weinstein--Aronszajn identity, we obtain
	\begin{equation}
		\log(\det\Sigma) = \sum_{\varyr, \varcntry} \log(\det\Sigma_{\varyr, \varcntry}) = \sum_{\varyr, \varcntry}  \left(\log(1 + n_{\varyr, \varcntry} \rho a^{-1}) + \log(a) n_{\varyr, \varcntry}\right).
	\end{equation}
	Using the Sherman–Morrison formula yields
	\begin{equation}
		r\tr \Sigma^{-1} r
		= 
		\sum_{\varyr, \varcntry} r_{\varyr, \varcntry}\tr \Sigma_{\varyr, \varcntry}^{-1} r_{\varyr, \varcntry}
		=
		\sum_{\varyr, \varcntry} \left( 
		a^{-1} \Vert r_{\varyr, \varcntry}\Vert^2 - \frac{\rho a^{-2}}{1 + n_{\varyr, \varcntry} \rho a^{-1}} (\mathbf{1}\tr r_{\varyr, \varcntry})^2
		\right),
	\end{equation}
	where $ r_{\varyr, \varcntry}\in\mathbb R^{n_{\varyr, \varcntry}}$ is the vector of residuals of the respective country and year and $\mathbf{1} := \begin{pmatrix} 1 \hdots 1 \end{pmatrix}\tr$.
	
	To fully comply with the methods associated to AIC and BIC, we would need to estimate the regression coefficients, $\sigma^2$, and $\rho$ by maximizing the new likelihood function. Unfortunately, the adaptation we made causes these estimates to be different from the common least squares approach, which is also used by KLW. This means that the coefficient value might change and the optimization problem might be difficult to solve. Thus, we use a simplified procedure by fixing the coefficients and $\sigma^2$ to the values of KLW's least squares regression and optimize only over $\rho$. 
	
\begin{figure}
		\includegraphics[width=\textwidth]{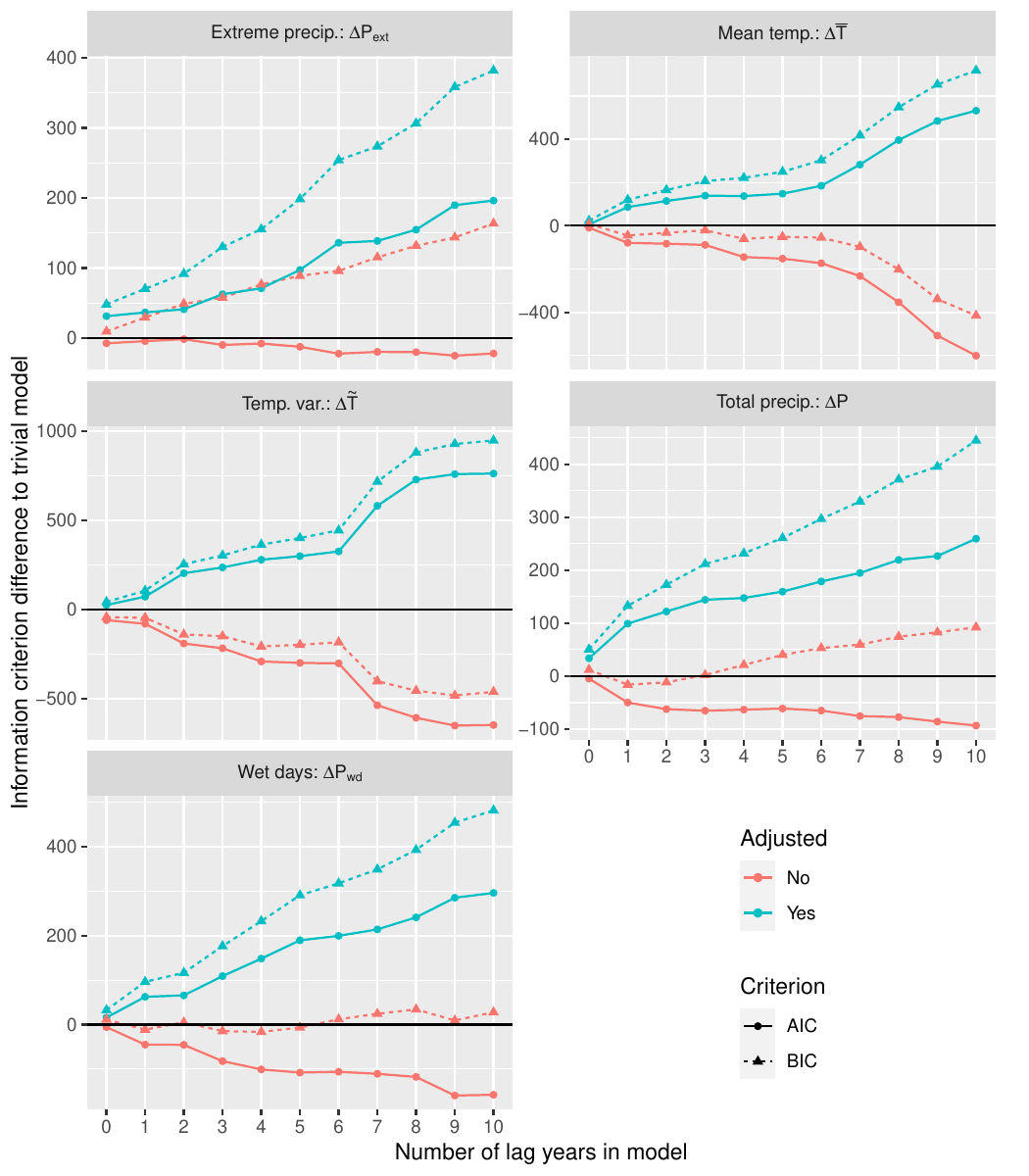}
		\caption{
			Change in adjusted and non-adjusted information criteria when adding a single term with lags to the trivial model.
			Minimal values show the preferred models.
		}\label{fig:icForward}
	\end{figure}

\begin{figure}
		\includegraphics[width=\textwidth]{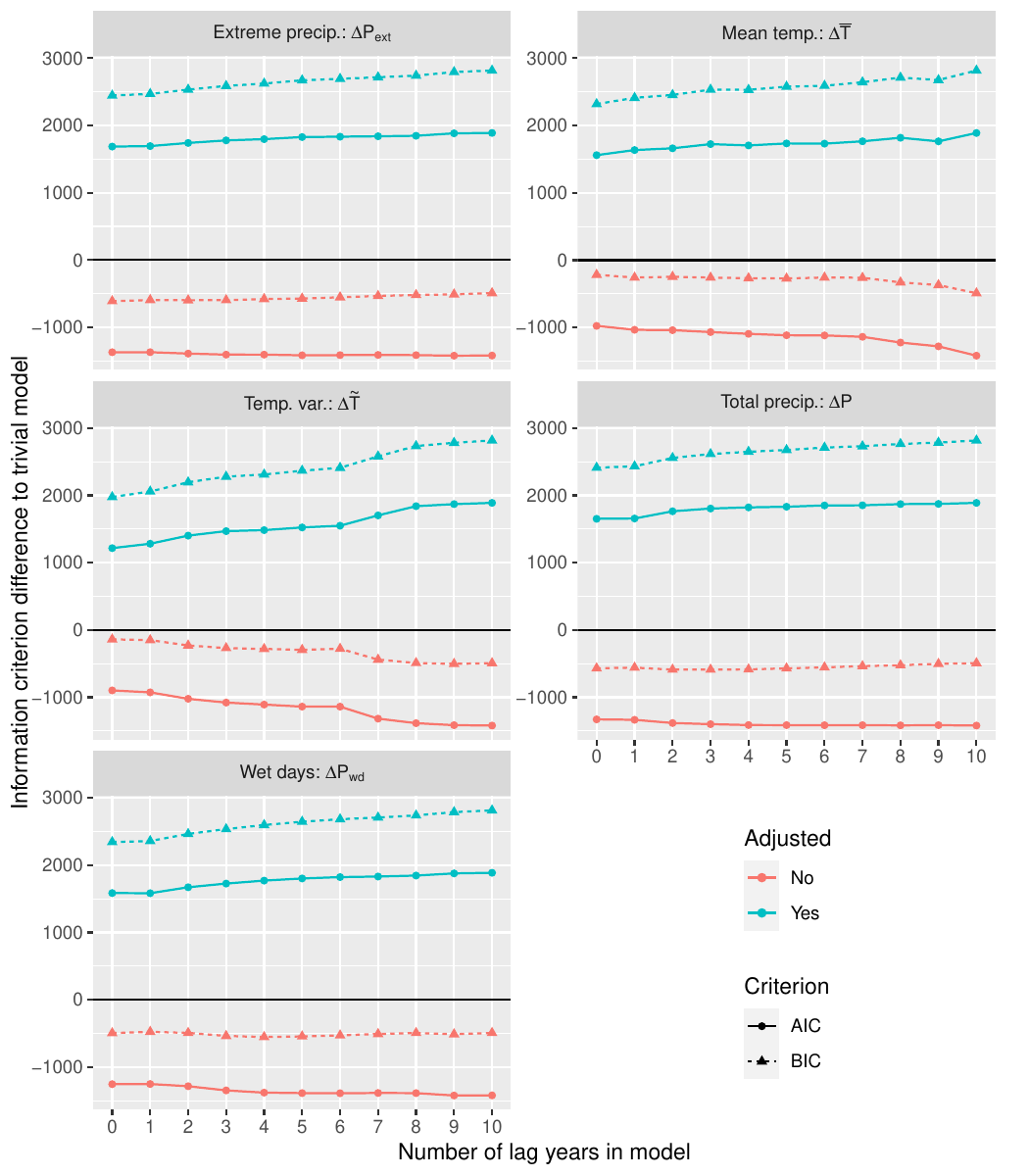}
		\caption{
			Change in adjusted and non-adjusted information criteria when removing lags of a single term from the full model with 10 lags per term. Minimal values show the preferred models.
		}\label{fig:icBackward}
	\end{figure}

	The results in \autoref{fig:icForward} and \autoref{fig:icBackward} discourage the use of any climate predictors when the correlation is adjusted for.
	
	\clearpage
	\section{Replications of KLW's Extended Data Fig.\ 1 with Different Clusterings}
    \label{sec:app:fig1}

	\begin{figure}
		\begin{center}
			\includegraphics[height=0.9\textheight]{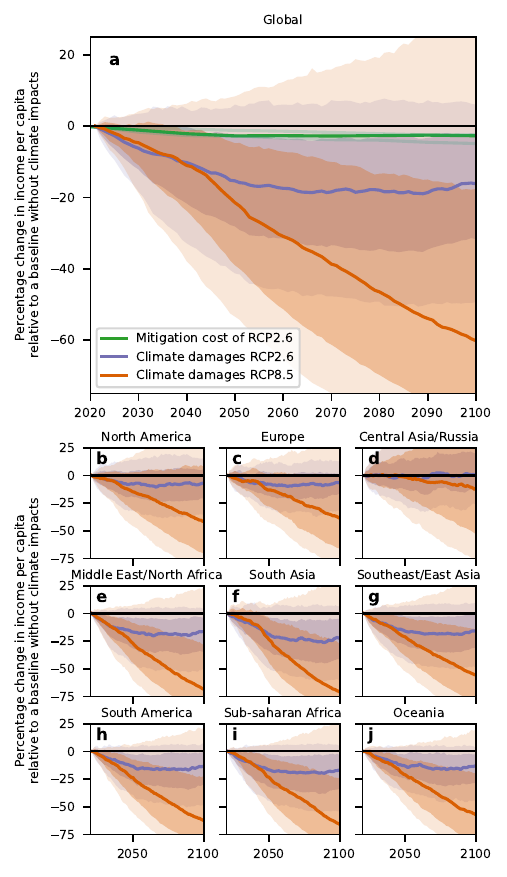}
		\end{center}
		\caption{
			As  Fig. 3 in the main text or KLW's Extended Data Fig.\ 1 but with clustering by country.
		}\label{fig:gid0}
	\end{figure}
	
	\begin{figure}
		\begin{center}
			\includegraphics[height=0.9\textheight]{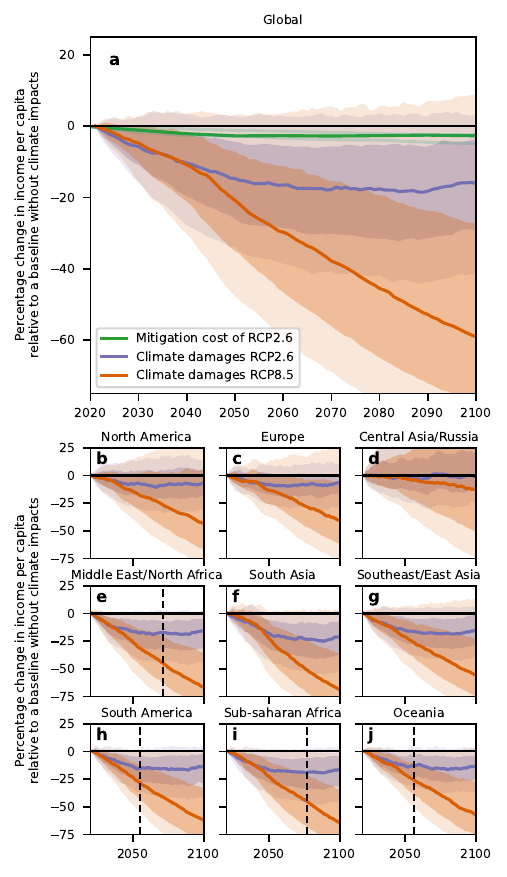}
		\end{center}
		\caption{
			As Fig. 3 in the main text or KLW's Extended Data Fig.\ 1 but with clustering by country--year.
		}\label{fig:gid0Year}
	\end{figure}
	
	\begin{figure}
		\begin{center}
			\includegraphics[height=0.9\textheight]{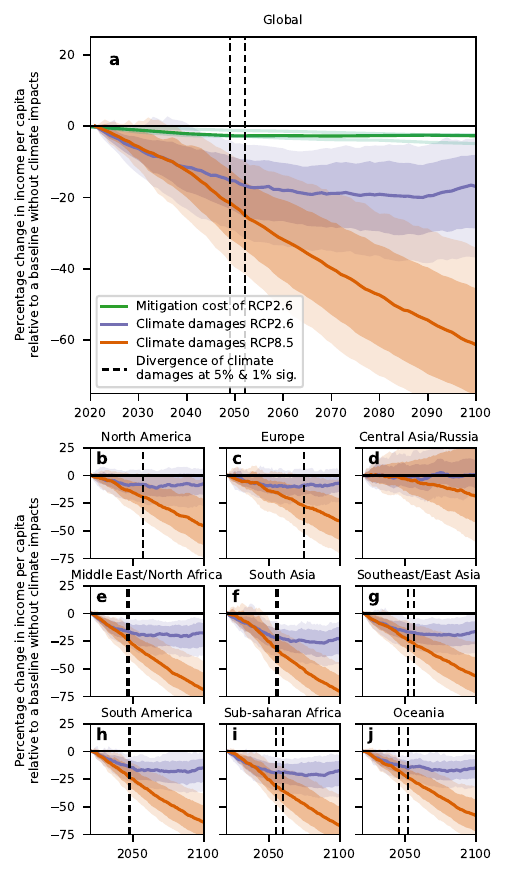}
		\end{center}
		\caption{
			As Fig. 3 in the main text or KLW's Extended Data Fig.\ 1 but with clustering by region.
		}\label{fig:gid1}
	\end{figure}
	
	\begin{figure}
		\begin{center}
			\includegraphics[height=0.9\textheight]{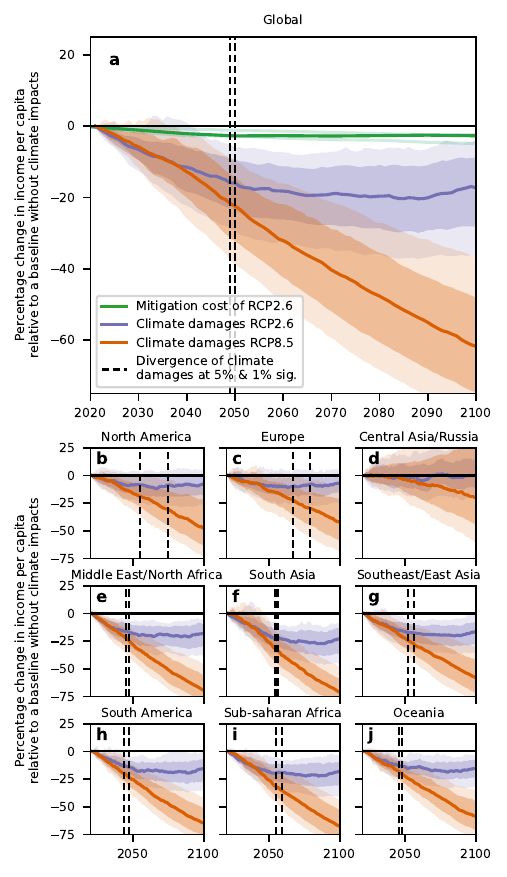}
		\end{center}
		\caption{
			As  Fig. 3 in the main text or KLW's Extended Data Fig.\ 1 but with clustering by region--year.
		}\label{fig:gid1Year}
	\end{figure}
\end{appendix}
\end{document}

%% file: packages.tex
\usepackage{amsthm,amsmath,amsfonts,amssymb}
\usepackage{mathtools}
\usepackage{stmaryrd}
\usepackage{enumitem} % advanced enumeration
\usepackage{graphicx} % for including graphics
\usepackage{dsfont} % indicator function \mathds{1}
\usepackage{tikz} % tikz pgf: advanced drawing

%% file: macros_general.tex
\newcommand{\R}{\mathbb{R}}

\newcommand{\sizedMid}[2]{#1 \, \kern-\nulldelimiterspace\mathopen{}\left| \vphantom{#1}\,#2\right.\mathclose{}\kern-\nulldelimiterspace}

\newcommand{\tr}{^{\!\top}\!} % transpose: A\tr